\documentclass{article}
\usepackage{spconf,amsmath,graphicx}
\usepackage{multirow,booktabs}   

\usepackage[noadjust]{cite}
\usepackage{hyperref}

\usepackage{enumitem}
\newenvironment{enumerate*}%
  {\begin{enumerate}%
    \setlength{\itemsep}{3pt}%
    \setlength{\parskip}{0pt}}%
  {\end{enumerate}}
\newenvironment{itemize*}%
  {\begin{itemize}%
    \setlength{\itemsep}{3pt}%
    \setlength{\parskip}{0pt}}%
  {\end{itemize}}

\newcommand\linesubsec[1]{\vspace{0.8mm}\noindent\textbf{#1 --- }}

\usepackage[dvipsnames]{xcolor}

\title{Automatic multitrack mixing with a differentiable\\ mixing console of neural audio effects}
\name{Christian J.~Steinmetz $^{1,2,\star}$ \qquad Jordi Pons $^1$ \qquad Santiago Pascual $^1$ \qquad Joan Serr\`a $^1$
\thanks{$^\star$ Work done during an internship at Dolby Laboratories.}}
\address{$^1$ Dolby Laboratories\\
$^2$ Music Technology Group, Universitat Pompeu Fabra
}
\begin{document}
\ninept
\maketitle
\begin{abstract}
Applications of deep learning to automatic multitrack mixing are largely unexplored. This is partly due to the limited available data, coupled with the fact that such data is relatively unstructured and variable. To address these challenges, we propose a domain-inspired model with a strong inductive bias for the mixing task. We achieve this with the application of pre-trained sub-networks and weight sharing, as well as with a sum/difference stereo loss function. The proposed model can be trained with a limited number of examples, is permutation invariant with respect to the input ordering, and places no limit on the number of input sources. Furthermore, it produces human-readable mixing parameters, allowing users to manually adjust or refine the  generated mix. Results from a perceptual evaluation involving audio engineers indicate that our approach generates mixes that outperform baseline approaches. To the best of our knowledge, this work demonstrates the first approach in learning multitrack mixing conventions from real-world data at the waveform level, without knowledge of the underlying mixing parameters.
\end{abstract}
\section{Introduction}
\label{sec:intro}

In the post-production process, the audio engineer is tasked with creating a cohesive mixture of the recorded elements. 
This process involves a number of technical challenges~\cite{case2011mix}, such as reducing masking, ensuring balance between the sources, and addressing noise or bleed, as well as artistic considerations, such as selecting the timbre and level of the artificial reverberation. 
Producing a mix is especially challenging due to the interplay between the aforementioned tasks, which are often dependant on each other, and not easily separated. 

Over the past decade, the accessibility of recording and music production equipment has rapidly increased. 
This, along with the growth and accessibility of digital distribution platforms, has brought music production to a new, diverse demographic~\cite{walzer2017bedroom}. 
Nevertheless, while these tools have become affordable and readily available, the skill and expertise required for their operation has remained relatively constant. 
Intelligent music production (IMP) is a research field focused on the development of algorithms that provide feedback, assistance, or automation in the process of recording, mixing, or mastering music~\cite{deman2019intelligent}. 
These methods often aim to address the high level of skill required in the music production process, lowering the barrier of entry, but their applications do not end there. 
Work in IMP may also help expedite the workflow of professional engineers, potentially uncover new understanding about current mixing conventions, or even discover new techniques for multitrack mixing. 

IMP systems generally implement either rule-based or classical machine learning approaches~\cite{moffat2019approaches}. 
Rule-based approaches rely upon establishing a set of rules and logic surrounding best practices~\cite{epg2009level, mansbridge2012ke, deman2013ke}.
While they generate convincing results for some cases, they do not provide a level of expressivity that matches human audio engineers~\cite{deman2017thesis}.
In comparison, classical machine learning approaches allow for greater model flexibility, but have typically suffered from the lack of parametric mixing data (i.e.,~the exact settings of each processor in the mix). For this reason, they have been of low-complexity, limiting their practical application~\cite{kolasinski2008ga, scott2011automatic, moffat2019drums}.
While both approaches have seen some success in addressing particular aspects of the mixing process, they ultimately have failed to capture the entire process and generalize at the scale of real-world projects. 

The previous shortcomings, along with the promise of deep learning methods in multiple audio signal processing tasks,
motivate the application of those within IMP. 
Nevertheless, there are a number of unique challenges in the application of deep learning methodologies to automatic mixing that have yet to be addressed: 
\begin{enumerate*}
    \item Large variation in the types and number of sources.
    \item Expectation for high-fidelity, requiring a low tolerance for artifacts and high sampling rates (at least 44.1\,kHz). 
    \item Ability for audio engineers to view and adjust the resulting mix parameters in an intuitive manner.
    \item Presence of many acceptable mixes and the difficulty in the objective evaluation of their quality.
\end{enumerate*}
In this paper, we address these challenges and demonstrate how we can successfully apply these methods.
Our major contributions are:
\begin{itemize*}
    \item We demonstrate that temporal convolutional networks can model a series connection of signal processing devices, across their parameter spaces. 
    \item We propose a differentiable mixing console enabling interpretability and the ability to learn from limited and unstructured data. 
    \item We introduce a loss function based on sum and difference signals, critical in enabling learning from real-world mixes.
    \item In a perceptual evaluation, we demonstrate that our model can learn mixing conventions from raw audio waveforms of real-world mixes, which we believe to be the first of its kind. 
\end{itemize*}

\begin{figure*}[ht] 
    \centering
    \includegraphics[width=\linewidth]{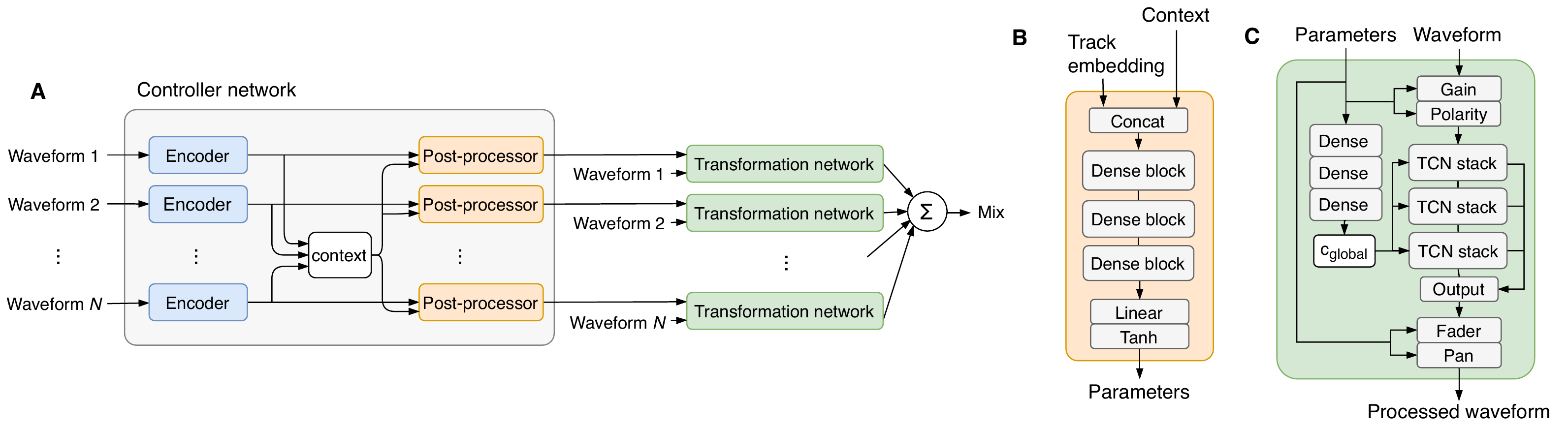} 
    \vspace{-0.6cm}
    \caption{
             Block diagrams of the DMC (A), its post-processor (B), and its transformation network (C).
             }
    \label{fig:complete-system}
\end{figure*}

\section{Differentiable mixing console}
\label{sec:dmc}

We aim to achieve a strong inductive bias for the mixing task by incorporating knowledge from the signal processing chain of a traditional mixing console. 
We consider a neural network that analyzes a set of audio inputs, and then predicts a set of parameters for each channel in the mixing console. 
In order to train this network, we need the ability to compute gradients through the processors in the channel (e.g.,~equalization, compression, and reverb).
This may be challenging, as implementations of these processors can be complex and varied. Additionally, they may not have easily-tractable or well-behaved gradients.
To overcome this, we propose to replace each channel in the mixing console with a neural network that aims to emulate, as closely as possible, the processing of the original channel, which we will call the transformation network.
We train this network by utilizing existing digital audio effects to generate training examples. 
We construct a differentiable mixing console (DMC), as shown in Fig.~\ref{fig:complete-system}, where each channel has been replaced by an instance of the pre-trained transformation network.
This enables us to train the controller network for the mixing task in an end-to-end fashion, without the need for the parameters used to create the ground truth mixes. 


\subsection{Transformation network}

\begin{figure*}[t]
    \centering
    \includegraphics[width=0.8\linewidth]{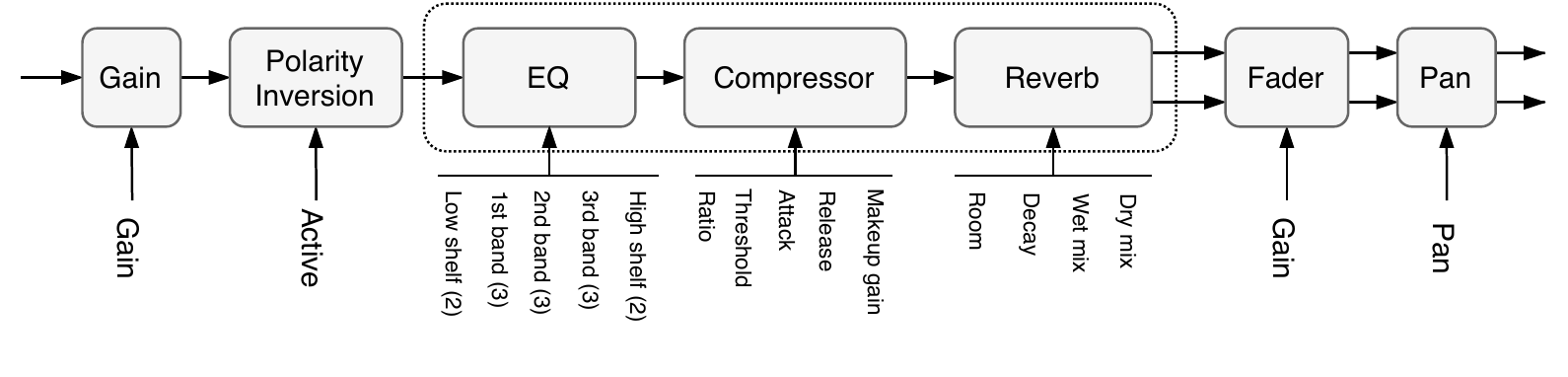}
    \vspace{-0.5cm}
    \caption{Block diagram of the signal processing chain in the differentiable mixing console channel. The labels indicate parameters passed to each processor. The three processors in the dashed box are modeled by the transformation network, while the others are implemented directly, since they pose no challenge in backpropagation.}
    \vspace{-0.25cm}
    \label{fig:channel-strip}
\end{figure*}

Various deep learning approaches have already been proposed for the task of modeling audio effects~\cite{covert2013rnn, zhang2018lstm, hawley2019compression, martinez2020blackbox, wright2020realj, wright2020perceptual}. 
While previous approaches have focused on training a single model for each effect, we believe our work is the first to consider building a model that emulates a series connection of effects and their parameters, jointly.
Most approaches do not consider modeling the different configurations of these devices, and those that do, only consider a sparse sampling of the parameters~\cite{hawley2019compression, wright2020realj}.  
This is due to the fact that they aim to emulate an analog device, and the process of collecting data at many configurations is often impractical.
However, we are interested in modeling the behavior of digital signal processors. 
As a result, we can generate effectively endless examples during the training of the transformation network. 
To this end, we developed a Python package, pymixconsole\footnote{\url{https://github.com/csteinmetz1/pymixconsole}}, which implements a framework for controlling a chain of audio effects found in a typical mixing console, as shown in Fig.~\ref{fig:channel-strip}. 
Using audio recordings of various musical sources at 44.1\,kHz from the SignalTrain dataset~\cite{hawley2019compression}, we generate training examples across all configurations of the chain, on-the-fly, by processing these recordings with uniformly sampled configurations. 


\begin{figure}[t]
    \centering
    \vspace{-0.0cm}
    \includegraphics[width=0.65\linewidth]{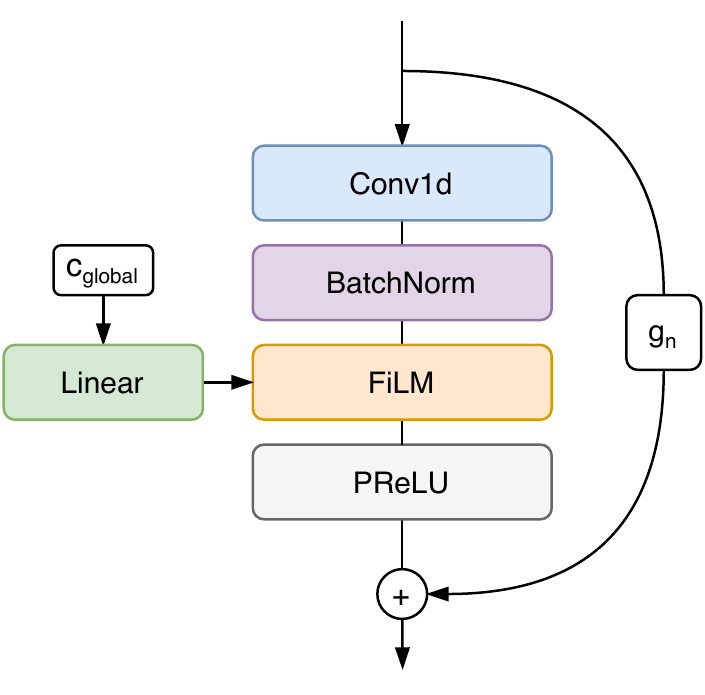}
    \vspace{-0.3cm}
    \caption{Block diagram of the TCN block.}
    \label{fig:tcn-block}
    \vspace{-0.3cm}
\end{figure}

For the design of the transformation network, we follow a similar implementation to Damskägg et al.~\cite{damskagg2019distortion} for modeling distortion effects, which adapts a non-causal WaveNet-like model~\cite{rethage2018wavenet}, formalized by Bai et al.~\cite{bai2018tcn} as the temporal convolutional network (TCN).
This network is composed of blocks of 1-dimensional convolutions, as shown in Fig.~\ref{fig:tcn-block}.
Exponentially increasing dilation factors and a kernel size of 15 are used to achieve a larger receptive field. 
We utilize batch normalization without an affine transformation, and couple it with feature-wise linear modulation (FiLM)~\cite{perez2018film} in order to inject conditioning information from the effect parameters.
The global conditioning $c_{\text{global}}$, shown in Fig.~\ref{fig:tcn-block}, 
is a vector generated by a small 3-layer multi-layer perceptron (MLP), which projects the signal chain parameters (e.g. 26 parameters for the complete channel) to a 128-dimensional vector. At each block, $c_{\text{global}}$ is projected via a linear layer to match the channel dimension for the FiLM operation.
A residual connection with a learnable gain is also included. 

We create multiple stacks of 10~TCN blocks to achieve a larger receptive field. 
The dilation factor $d$ of layer $l$ is given by $d_l = 2^{(l - 1) \mod 10}$, where $\mod$ is the modulo operation. 
We consider three configurations of the TCN with 10, 20, and 30~blocks each: TCN-10, TCN-20, and TCN-30, which achieve receptive fields of 320\,ms, 650\,ms and 970\,ms, respectively.
Additionally, skip connections are included from the intermediate activations from every layer, where they are averaged before being added to the final layer.

Since the channel includes processing operations that are differenatiable, like the input gain, polarity, fader, and panning, we implement these directly in the transformation network, before and after the TCN stacks, as shown in Fig.~\ref{fig:complete-system}C. The TCN is trained to emulate only the more complex processors in the channel, shown in the dashed box in Fig.~\ref{fig:channel-strip}: the EQ, compressor, and reverb. With this configuration, we will define two different mixing tasks. First, a basic mix, where the TCN is removed and only the gain and panning are predicted by the controller, and second, a full mix, where the TCN is active, and parameters for all the processors are predicted by the controller to fully emulate the processing in the mixing console.


\subsection{Controller network}

The controller network contains a series of encoders and post-processors with shared weights, as shown in Fig.~\ref{fig:complete-system}A. 
First, the encoder must learn to extract and aggregate information from the inputs that is salient for the mixing process.
The encoder is constructed following the common spectrogram-based VGGish model~\cite{hershey2017cnn}, 
and we conduct transfer learning by using the pre-trained weights on AudioSet provided by Gemmeke et al.~\cite{gemmeke2017audioset}.
We found that fine-tuning these weights during training further improved performance. 

The post-processor, shown in Fig.~\ref{fig:complete-system}B, is a simple 3-layer MLP, with PReLU activations and dropout of $0.1$.
Recall that the weights of both the controller and transformation networks are shared across all input channels. 
This means the process of predicting parameters for each channel occurs independently on a per-track basis.
Therefore, by default, cross-channel interactions cannot be captured, a critical consideration for creating a mix~\cite{vega2010quantifying}. 
To address this issue, each instance of the post-processor is passed two inputs, the track embedding for the respective input channel, and an additional context embedding, which is computed by simply averaging the embeddings generated for all of the input sources. 
While this may obscure some of the information about the input sources in the mix, we found this provided sufficient context and worked in practice. 

\subsection{Stereo loss function}

\begin{figure*}[t]
    \vspace{-0.3cm}
    \begin{minipage}[c]{0.478\linewidth}
        \centering
        \includegraphics[width=\linewidth]{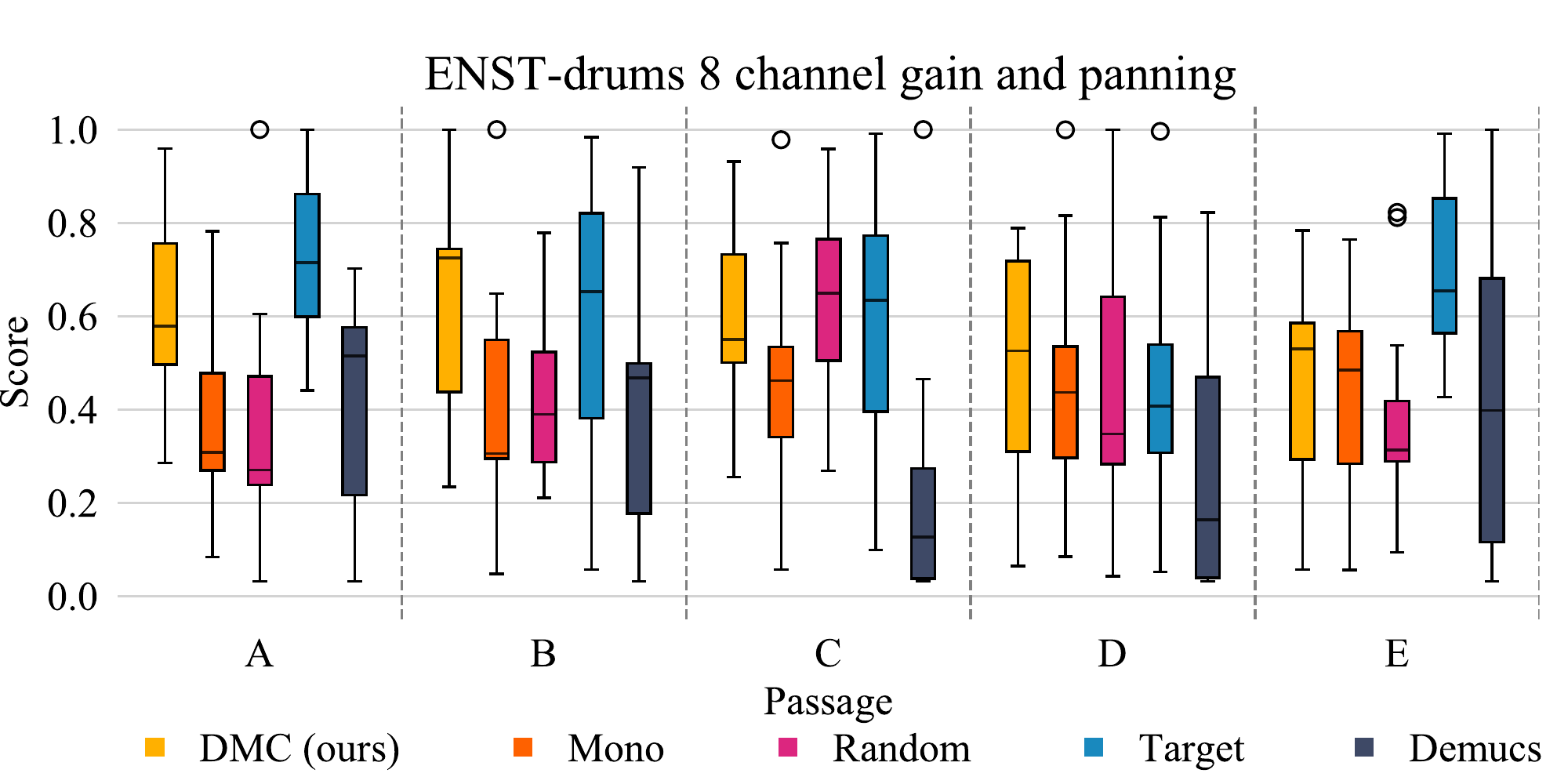}
    \end{minipage}
    \hfill
    \begin{minipage}[c]{0.478\linewidth}
        \centering
        \includegraphics[width=\linewidth]{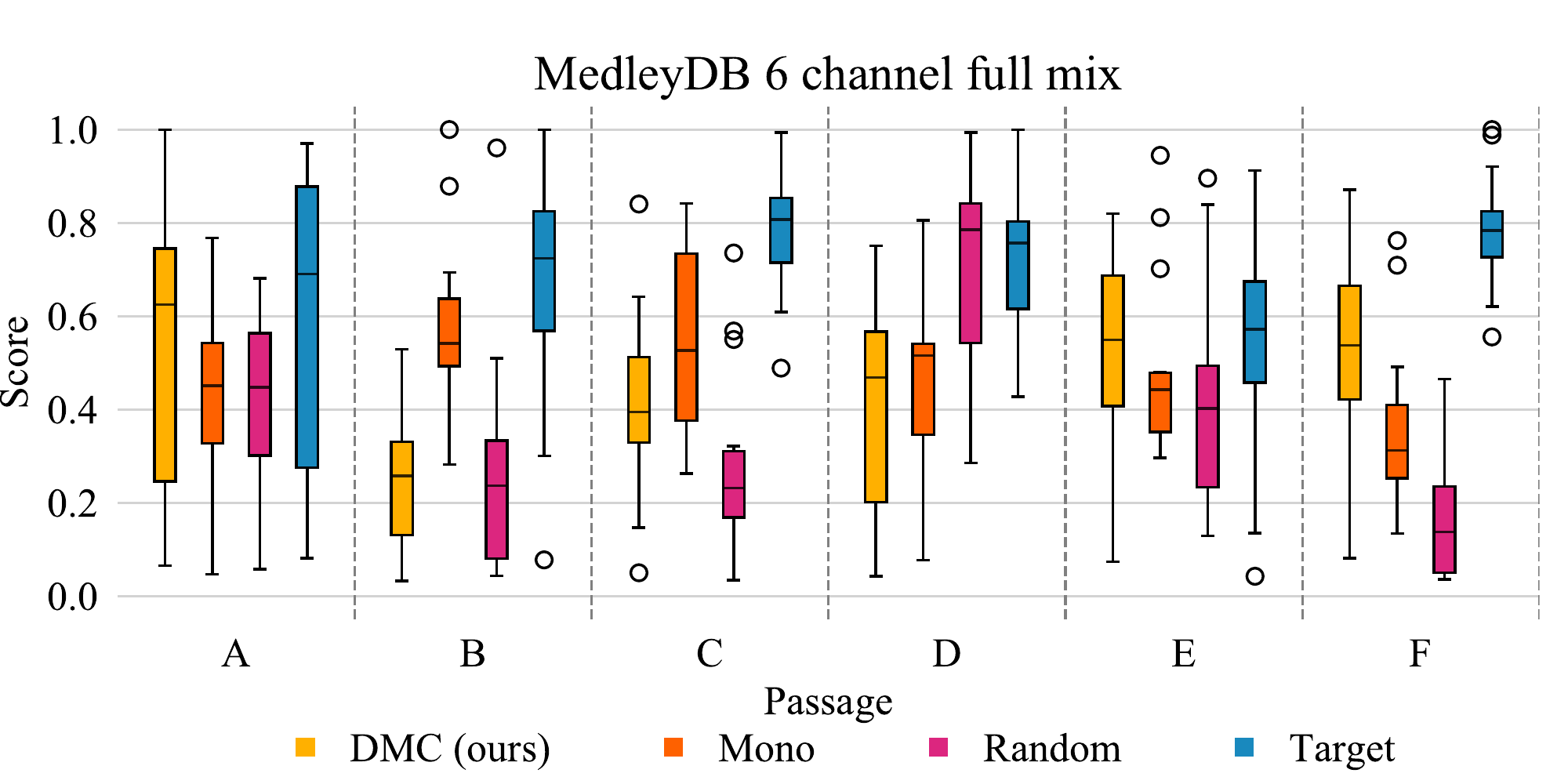}
    \end{minipage}%
    \vspace{-0.2cm}
    \caption{Perceptual evaluation on the ENST-Drums gain and panning mixing task (left), and for MedleyDB full-channel mixing task (right).}
    \label{fig:evaluation}
    \vspace{-0.4cm}
\end{figure*}

A critical step in the mixing process involves panning the elements in the mix across the stereo field. 
This must be done in such a way to achieve proper balance between the left and right channels, while also ensuring the appropriate spatialization of the elements. 
When training a model to generate stereo mixes using the L1 loss or multi-resolution STFT loss~\cite{yamamoto2020parallel}, there is an inherent issue: these loss functions applied to multi-channel audio heavily penalize mixes that place elements on the opposite side of the stereo field compared to the ground truth mix. 
This poses a challenge, since these loss functions encourage the model to always place sources in the center of the stereo field.
Consider a ground truth mix with an electric guitar panned to the left. 
While the model may know that the guitar ought to be panned to the left or the right side and not the center, it has no way of predicting the absolute orientation in the ground truth mix.
Therefore, to minimize the error, the model is incentivized to place it in the center, creating a more perceptually inaccurate mix. 


To address this issue, we design a stereo loss function that aims to achieve left-right invariance, so only the overall stereo balance is considered. We first compute the sum and difference signals,
\begin{align*}
    y_{\text{sum}}  &= y_{\text{left}} + y_{\text{right}}   \\
    y_{\text{diff}} &= y_{\text{left}} - y_{\text{right}}  ,
\end{align*}
directly on the left and right channels of the time-domain signals. While this does transform the stereo information to another representation, the absolute stereo orientation is now represented in the phase of the difference signal. To ignore this phase information, we apply a multi-resolution STFT loss~\cite{yamamoto2020parallel}, which is composed of a spectral convergence (SC) and a spectral log-magnitude (SM) term:
\begin{align*} \label{eq:spectral-convergence}
    \ell_{\text{SC}}({\hat{y}},y) &= \frac{\| ~|\text{STFT}(y)| - |\text{STFT}(\hat{y})|~ \|_{\text{F}}}{\| ~|\text{STFT}(y)|~ \|_{\text{F}}} \\ 
    \ell_{\text{SM}}({\hat{y}},y) &= \frac{1}{N} \left\| \log \left( |\text{STFT}(y)| \right) - \log \left( |\text{STFT}(\hat{y})| \right) \right\|_1,
\end{align*}
where $|\text{STFT}(\cdot)|$ is the short-time Fourier transform magnitude, $||\cdot||_{\text{F}}$ is the Frobenius norm, $||\cdot||_1$ is the L1 norm, $N$ is the number of STFT frames, and $\hat{y}$ and $y$ denote the predicted and target signals, respectively. 
To compute the final multi-resolution STFT loss, $M$ different STFT configurations are chosen with varying window, hop, and frame sizes, and the error at these resolutions is averaged: 
\begin{equation*} 
     \ell_{\text{MR}}({\hat{y}},y) = \frac{1}{M} \sum_{m=1}^{M} \left( \ell_{\text{SC}}({\hat{y}},y) + \ell_{\text{SM}}({\hat{y}},y) \right).
\end{equation*}
In all our experiments, we follow the original multi-resolution STFT implementation proposed by Yamamoto et al.~\cite{yamamoto2020parallel}, which includes three different STFT frame sizes of $512$, $1024$, and $2048$. 
Applying $\ell_{\text{MR}}$ as defined above to both the sum and difference signals, we define the stereo loss function
\begin{equation*}
    \ell({\hat{y}},y) = \ell_{\text{MR}}(\hat{y}_{\text{sum}},y_{\text{sum}}) + 
    \ell_{\text{MR}}(\hat{y}_{\text{diff}},y_{\text{diff}}).
\end{equation*}

\subsection{Training}

We begin by training the transformation network by minimizing the mean absolute error (MAE), or L1 loss, between the predicted and ground truth processed waveforms.
Since the predicted waveforms are smaller than the input waveform due to the lack of padding in the model, we take a central crop from the ground truth that matches the size of the predicted waveform. 
We use Adam, a learning rate of $3\cdot 10^{-4}$, and a batch size of 32, 
along with plateau learning rate scheduling, halving the learning rate after the validation loss has not decreased for 20~epochs, defining an epoch as 1,000~random 1.5-second patches from the dataset.

To train the DMC, we create instances of the pre-trained enocder and transformation network, along with the post-processor, for each input source in the multitrack input. 
We again use Adam and the same learning rate, 
but we scale the batch size due to memory constraints: for the basic task we use a batch size of 16 and for the full task we use a batch size of 2.
For all DMC models, we again use plateau learning rate scheduling, halving the learning rate after the validation loss has not decreased for 200~epochs, and defining an epoch as 100~random 5-second patches.

\section{Evaluation and Results}
\label{sec:Results}

\subsection{Audio effect modeling} 

We evaluated variants of the TCN for channel emulation task by comparing the MAE, as well as the multi-resolution STFT distance, between the ground truth and predicted waveforms.
We found that increasing the receptive field improved performance, with the TCN-30 (MAE: 0.024, STFT: 2.210) outperforming the shallower models, TCN-20 (MAE: 0.027, STFT: 2.315) and TCN-10 (MAE: 0.035, STFT: 2.701)
Additionally, on a separate task of modeling an analog compressor, our TCN-20 (MAE: $4\cdot 10^{-3}$, STFT: 0.606) outperformed the current state-of-the-art for this task, SignalTrain~\cite{hawley2019compression} (MAE: $8\cdot 10^{-3}$, STFT: 1.657), by a substantial margin.
Further details of these evaluations are presented in~\cite{steinmetz2020msc} (Ch.~4).

\subsection{Multitrack mixing}

\linesubsec{Datasets}
First, we consider the ENST-Drums dataset~\cite{gillet2006enst}, which includes around 3\,h of multitrack recordings of drummers.
Each example in the dataset includes 8 sources from the drum kit. 
We generate training, validation, and test splits (80/10/10) following Moffat et al.~\cite{moffat2019drums}. 
This dataset provides an initial indication of the ability to learn mixing conventions from mixes with consistent sources and mixing techniques across the dataset. 
Next, to demonstrate the ability of our framework to generalize to real-world use cases, we further consider MedleyDB~\cite{bittner2014medleydb, bittner2016medleydb}, which provides realistic and diverse multitrack recordings of complete songs across a number of genres. 
The dataset contains 196~songs with around 7\,h of recordings. 
Due to memory constraints, we train models using songs with $\leq$ 16~inputs for the basic task and $\leq$ 6~inputs for the full task, resulting in 120 and 65 songs, respectively. 
Similarly to ENST-Drums, we create an 80/10/10 split of the data, plus we ensure that songs from the same artist do not fall into different splits. 

\linesubsec{Baselines}
We consider three baselines. 
The first one is the mono mix, a sum of the inputs. 
The second one scales each input by a random gain ($-12$\,dB to $+12$\,dB), along with random panning.
The third one is what we consider a canonical deep learning approach for processing time domain signals: we adapt the Demucs architecture~\cite{defossez2019music}, originally designed for source separation. Unlike our DMC model, this architecture does not present an inductive bias for the mixing task. 
To be comparable to the DMC model, we remove the LSTM layers from the center of the original network, and also scale down the number of channels, which results in around 80\,M parameters.
We pass a fixed number of input sources and train the model to predict the ground truth mix from the dataset. 
For songs with fewer inputs than we train with, we fill these inputs with zeros.

\linesubsec{Perceptual evaluation}
Due to the subjective nature of mixing, we conduct a perceptual evaluation with the Web Audio Evaluation Tool~\cite{jillings2016weat} using the APE test design \cite{de2014ape}.
We enlisted 16~audio engineers with mixing experience.
They were presented passages from the test set mixed by each approach, and were instructed to rate each on a scale from 0 to 1.
More details are shown in~\cite{steinmetz2020msc} (Ch.~5).

\linesubsec{Results}
We first report results on the ENST-Drums dataset, on the basic mixing task in Fig.~\ref{fig:evaluation} (left).
On average, the target mixes tend to be rated the highest, with mixes from our DMC following close behind, and even surpassing the target mixes in the case of passages B and D. 
On average, the mono and random mixes were rated lower, with the Demucs-like model being constantly rated the lowest (listeners indicated there were artifacts, likely from the transposed convolutions used by the model). 
To formalize these results, we perform the Kruskal-Wallis H-test, which points to a difference between the approaches ($F=64.01$, $p=8\cdot 10^{-14}$). 
Our further ad-hoc analysis with Conover's test reveals there is not a significant difference between the target and DMC mixes ($p_{\text{adj}}=0.08$).

We continue with MedleyDB and the full mixing task. 
We omit the Demucs-like model since it was unable to converge when training on MedleyDB, which we posit was a result of its lack of permutation invariance. 
Results are shown in Fig.~\ref{fig:evaluation} (right).
The Kruskal-Wallis H-test again reveals that there is a difference between the approaches ($F=48.1$, $p=8.8\cdot 10^{-10}$), and Conover's test reveals that all approaches perform differently from each other, with $p_{adj}=1.1\cdot 10^{-9}$ for the target vs.\ DMC comparison. 
Interestingly, in passages A and E, the DMC is nearly on par with the target. However, in B, C, and D, DMC performs poorly. In F, DMC performs clearly better than the baselines, but does not reach the level of the target. Note also, that random mixes include only gain and panning, and not the entire signal chain. Therefore, the DMC has a much more challenging task. Our listening suggests that failure cases arise from the over application of reverb on elements like the vocals, which listeners rated harshly. We provide listening examples in  \url{https://csteinmetz1.github.io/dmc-icassp2021/}. 

\section{Discussion}
\label{sec:discussion}

We outline and address a number of challenges in applying deep learning methods to build a model for automatic mixing trained directly on realistic multitrack data. We build a model with a strong inductive bias for this task, taking inspiration from the mixing console. By employing pre-trained sub-networks, weight sharing, and a stereo loss function, we demonstrate, to our knowledge for the first time, the ability to learn mixing conventions directly from waveforms of real-world multitracks. In the process, we demonstrate that the TCN can adequately model a series connection of effects over a dense sampling of their parameters. While the results on the complete mixing task are somewhat limited, we hypothesize that with larger models and more data, performance could be further improved. 




\bibliographystyle{IEEEbib}
\bibliography{strings,references}

\end{document}